\documentclass[aps,prl,twocolumn,amsmath,superscriptaddress,showpacs]{revtex4}

\newif\ifpdf\ifx\pdfichsageja\undefined\pdffalse\else\pdftrue\fi
\ifpdf\pdfoutput=1\usepackage[pdftex]{graphicx}\usepackage{epstopdf}\pdfcompresslevel=9
\else\usepackage{graphicx}\fi

\usepackage{graphicx}
\usepackage{amssymb}
\usepackage{amsmath}
\usepackage{epsfig}

\begin{document}

\title{Chaos Synchronization with Dynamic Filters: Two Way is Better Than One Way}

\author{Ido Kanter}
\affiliation{Department of Physics, Bar-Ilan University, Ramat-Gan, 52900 Israel}
\author{Evi Kopelowitz}
\affiliation{Department of Physics, Bar-Ilan University, Ramat-Gan, 52900 Israel}
\author{Johannes Kestler}
\affiliation{Institute for Theoretical Physics, University of W\"urzburg, Am Hubland, 97074 W\"urzburg, Germany}
\author{Wolfgang Kinzel}
\affiliation{Institute for Theoretical Physics, University of W\"urzburg, Am Hubland, 97074 W\"urzburg, Germany}

\begin{abstract}
Two chaotic systems which interact by mutually exchanging a signal
built from their delayed internal variables, can synchronize.  A third
unit may be able to record and to manipulate the exchanged signal. Can the third unit
synchronize to the common chaotic trajectory, as well? If all
parameters of the system are public, a proof is given that the
recording system can synchronize as well.  However, if the two
interacting systems use private commutative filters 
to generate the
exchanged signal, a driven system cannot synchronize.  It is shown
that with \emph{dynamic} private filters the chaotic trajectory even cannot be
calculated. Hence two way (interaction) is more than one way (drive).
The implication of this general result to secret communication with
chaos synchronization is discussed.
\end{abstract}


\maketitle


Chaos synchronization is one of the most intriguing phenomena in the
wide field of synchronization. On one hand a chaotic system is very
unpredictable, and two chaotic systems, starting from almost
identical initial states, end in completely uncorrelated
trajectories \cite{1}. On the other hand, two chaotic systems which
are coupled by some of their internal variables can synchronize to a
common identical chaotic motion \cite{2,3}. This phenomenon has
attracted a lot of attention, mainly because of its potential for
secure communication. A secret message 
can be modulated on the
chaotic signal of a sender, and a receiver with an identical system
which is driven by the modulated signal can decrypt this
message \cite{CPF1,CPF2}. In fact, communication with chaos
synchronization has recently been demonstrated with semiconductor
lasers which were synchronized over a distance of $120$ km in a
public fiber network \cite{nature}.

Such a unidirectional configuration, a sender A is driving a
receiver B, is susceptible to an attack. A third unit E, which is
coupled to the transmitted signal, can synchronize as well, provided
it has identical parameters. Therefore, a bi-directional
configuration has been suggested where two chaotic units interact
and synchronize by their mutual signals \cite{lasers2_MCPF,Einat1}. 
In this case, a driven unit
E responds differently to the signal than the two interacting
partners A and B. This opens the possibility for public encryption
protocols. Although an attacker knows all parameters of the system
and although he can record any transmitted signal, he cannot decrypt
the secret message.

This kind of public encryption relies on the fact, that two chaotic
systems A and B synchronize by bi-directional interaction whereas a
third unit E which is only driven by the transmitted signal cannot
synchronize. However, it is not obvious that this is possible, at
all. On one hand, the two mutually coupled chaotic systems
influence the dynamics of each other and can accelerate the
synchronization by enhancing coherent moves, whereas the
unidirectionally coupled system, a listener, cannot influence the
synchronization process. On the other hand, the listener is allowed
to record the exchanged signals and to manipulate the recorded signals,
without affecting the synchronization
process \cite{MRZ}. In fact, Taken's theorem provides a
mathematical proof that it is possible to reconstruct the complete
chaotic trajectory from the transmitted signal
\cite{Pikovsky_private_communication}. Hence, in principle an
attacker  
may be able to calculate the chaotic trajectory. 
In practice, 
however,
it may be difficult to achieve
perfect synchronization, in particular for
realizations with semiconductor lasers \cite{lasers2_MCPF,lasers1}. But the main
problem 
of nonlinear dynamics
remains: Is it possible that two interacting chaotic units
synchronize whereas a third unit which is driven by the transmitted
signal cannot synchronize? Note that the two partners are not
allowed to exchange any secret information; the attacker E knows all
the details which A knows about the system of B and vice versa.

The question raised before -- in short "Is two-way better than
one-way?" -- is addressed in this Letter. First, for identical
partners which synchronize by a bi-directional signal we prove that
an attacking chaotic unit can synchronize, as well. However, for
non-identical partners which use private commutative filters we show
that two interacting units can synchronize whereas a driven unit can
neither synchronize nor calculate the synchronized trajectory from
the transmitted signal.

We differentiate between two possible types of unidirectional
listeners: hardware listeners and software listeners. A hardware
listener consists of a similar chaotic setup to those of the
synchronized chaotic partners, whereas a software listener is
capable to record and to mathematically manipulate the 
recorded signal.


We start with a proof that a hardware listener is possible for any
identical chaotic units which interact by exchanging a function of
some of their internal variables. This function may be nonlinear and
may contain delayed variables. For simplicity, we consider the
simple case of iterated chaotic maps, but the proof holds for
ordinary differential equations with delay, as well. Consider two
partners $x^A$ and $x^B$. Their dynamics is controlled by a general
self-feedback function $f$ and a general coupling function $g$ which
are both nonlinear functions of the history $\tau$ steps back
\begin{equation}
\begin{split}
x_t^A=f(\vec{x}^A_t) + g(\vec{x}^B_t) \\ 
x^B_t=f(\vec{x}^B_t) + g(\vec{x}^A_t)
\end{split}
\label{eqn_1}
\end{equation}
where $\vec{x}_t=(x_{t-1},..,x_{t-\tau})$. Since the
coupling is symmetric, 
these equations may be transformed to the
center of mass of the partners, $d_t =
\frac{1}{2}(x^A_t - x^B_t)$ and $s_t = \frac{1}{2}(x^A_t +
x^B_t)$,
which gives
\begin{eqnarray}
2d_t = f(\vec{s}_t + \vec{d}_t) - f(\vec{s}_t - \vec{d}_t) +
g(\vec{s}_t - \vec{d}_t) - g(\vec{s}_t + \vec{d}_t)
\label{eqn_2dt}
\end{eqnarray}
where $2d_t$ is actually the distance between $x^A$ and
$x^B$ and is also a measure of their synchronization. Linear
expansion for small $d_t$ gives
\begin{eqnarray}
d_t = \nabla_{\vec{s}_t}f\cdot \vec{d}_t -
\nabla_{\vec{s}_t}g\cdot \vec{d}_t \label{eqn_dt} .
\end{eqnarray}
The dynamics of the listener, $x^E_t$, is influenced by both
transmitted signals from $x^A$ and $x^B$, and he compensates the
two received signals by subtracting the same amount from its own
signal
\begin{equation}
x^E_t = f(\vec{x}^E_t) - g(\vec{x}^E_t) + g(\vec{x}^A_t) +
g(\vec{x}^B_t) \label{eqn_et} .
\end{equation}
The distance between the listener and the partner $x^A$, for
instance, is defined by $e_t=x^A_t-x^E_t$. Applying the same
transformation as above and linear expansion for small $e_t$ give
\begin{equation}
e_t =\nabla_{\vec{s}_t+\vec{d}_t}f\cdot \vec{e}_t-
\nabla_{\vec{s}_t+\vec{d}_t}g\cdot \vec{e}_t \label{eqn_ct} .
\end{equation}
The comparison between (\ref{eqn_dt}) and (\ref{eqn_ct})
indicates that both dynamics are governed by the same conditional
Lyapunov exponents, which implies that the listener will
synchronize together with the partners. Hence, there is no
advantage in mutual coupling over unidirectional coupling when
dealing with transmitted signals which are an identical public
function of the output of the deterministic chaotic maps.

The hardware listener, defined by Eq.~(\ref{eqn_et}), is able to synchronize to
the two interacting units A and B, a software listener is not
required for this case.


But now we extend the configuration, Eq.~(\ref{eqn_1}), to the case of
non-identical units $x^A$ and $x^B$. Both units are using
different functions $g_A$ and $g_B$, and the two transmitted
signals are $g_B(\vec{x}^B_t)$ and $g_A(\vec{x}^A_t)$. These
functions are private, only $x^A$ knows $g_A$ and $x^B$ knows
$g_B$, and they commute, $g_A(g_B(x)) = g_B(g_A(x))$. Since a
listener does not know these functions, he cannot use them for his
hardware attack. On the other side, when the two partners $x^A$
and $x^B$ are synchronized, $x^A_t=x^B_t$, they receive an
identical drive $g_A(g_B(\vec{x}^A_t))$. In the following we use
linear filters for the two functions $g$. Do the two chaotic units
$x^A$ and $x^B$ synchronize in this case?

\begin{figure}[h]
\includegraphics[scale=0.7]{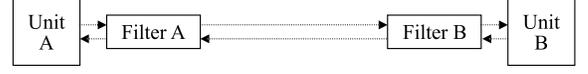}
\caption{\label{fil_1} A setup of two time-delayed, mutually coupled
units, where each unit has a filter influencing both transmitting
and receiving signals.}
\end{figure}

\begin{figure}[h]
\includegraphics[width=0.235\textwidth,height=0.22\textwidth]{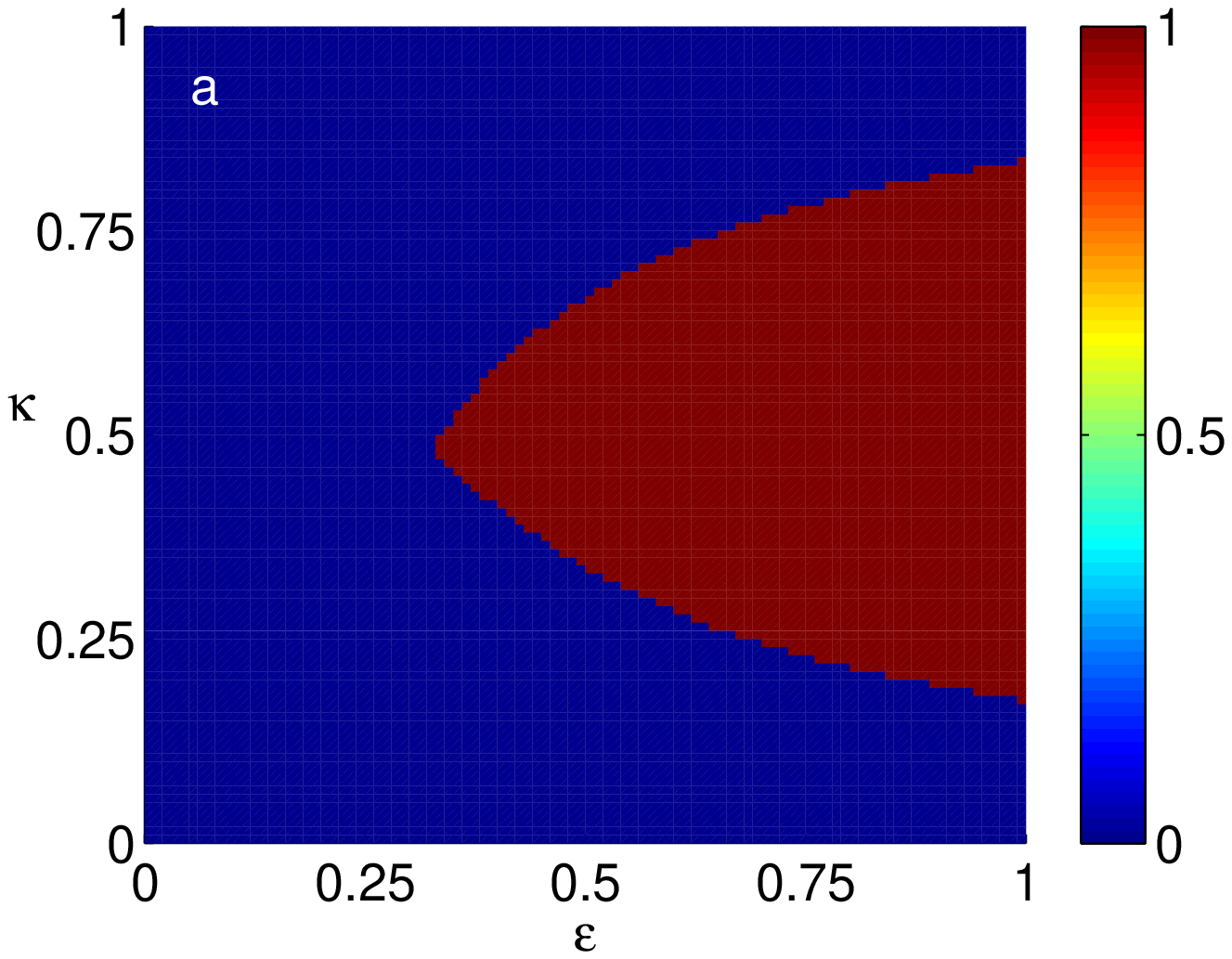}
\includegraphics[width=0.235\textwidth,height=0.22\textwidth]{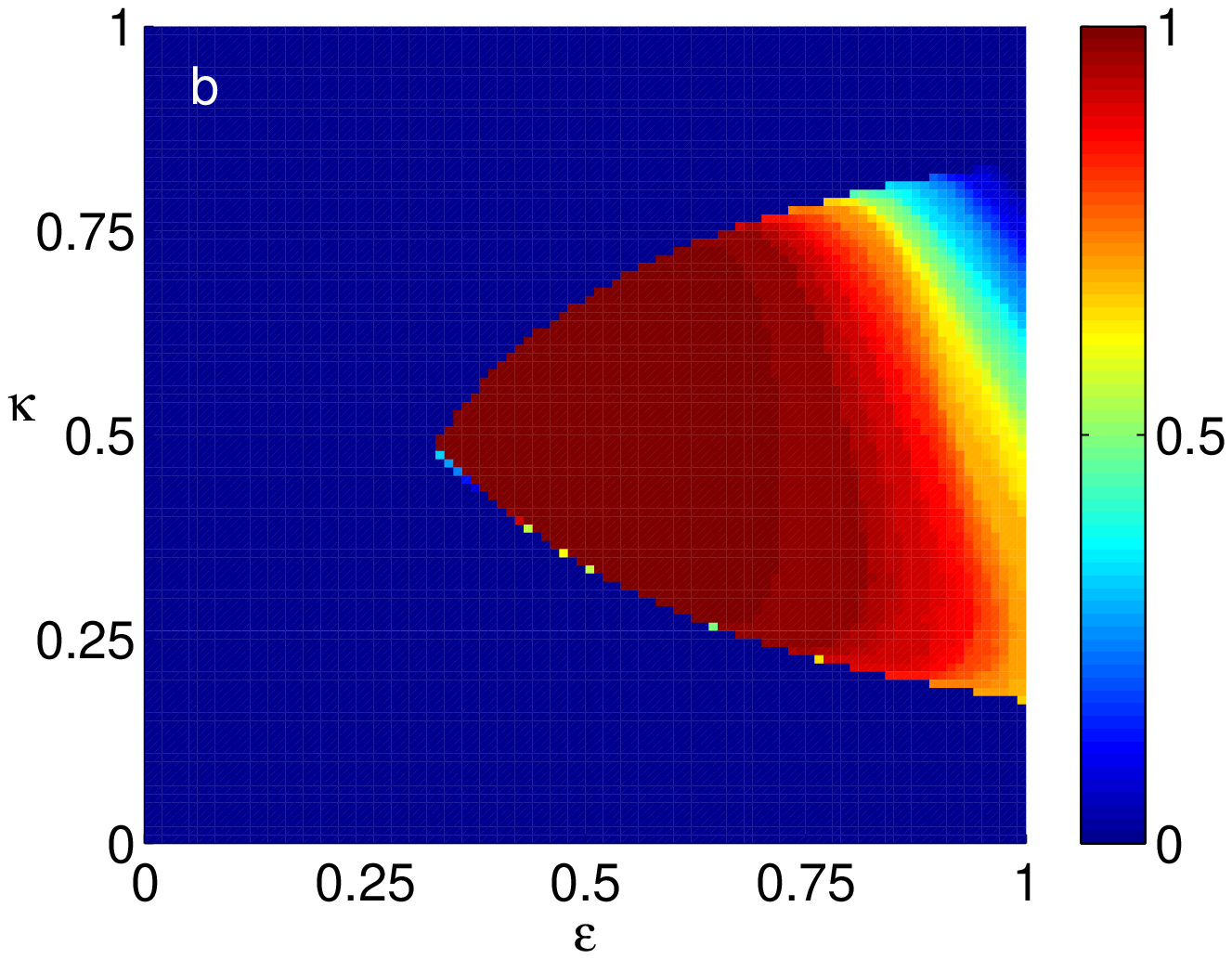}
\caption{\label{fil_2} Analytic results for the fraction of the phase
space, $(\varepsilon,\kappa)$, where synchronization is achieved
for a Bernoulli map with $\tau=40$ and $a=1.5$. (a) With the
absence of filters, synchronization is achieved only in the red
regime. (b) The probability to synchronize in the  phase space in
the case of a static filters with $N=2$.}
\end{figure}

This 
question is answered
below by one of the simplest chaotic
maps, the Bernoulli map, and with 
linear filters. 
Without filters, 
the dynamics
of the two mutually coupled units $x_t^A$ and $x_t^B$ can
be analyzed analytically and is given by
\begin{equation}
\begin{split}
x_{t+1}^A=(1-\varepsilon)f(x_t^A)+\varepsilon[\kappa
f(x_{t-\tau}^A)+(1-\kappa)f(x_{t-\tau}^B)] \\
x_{t+1}^B=(1-\varepsilon)f(x_t^B)+\varepsilon[\kappa
f(x_{t-\tau}^B)+(1-\kappa)f(x_{t-\tau}^A)]
\end{split}
\end{equation}
where $f(x)=(a x) \mod 1$.
The parameter
$\varepsilon$ indicates
the weight of the delayed terms, $\kappa$ stands for the strength of
the self-coupling term and a Bernoulli map is chaotic for $a>1$.
Note that $[0,1]$ is the allowed range for $\varepsilon$ and
$\kappa$. A linear expansion of the distance $d_t=x_t^A-x_t^B$ 
leads to
\begin{equation}
d_{t+1}=(1-\varepsilon)ad_t+\varepsilon a(2\kappa-1)d_{t-\tau} .
\end{equation}
By assuming that the distance converges/diverges exponentially in
time, $d_t=c^t$, it is possible to find an expression for the largest
conditional Lyapunov exponent \cite{physica_d,joha}.
We find that the largest Lyapunov exponent is negative and
synchronization is achieved for
\begin{equation}\frac{a-1}{2a\varepsilon}<\kappa<\frac{2a\varepsilon+1-a}{2a\varepsilon}\end{equation}
as is depicted in figure \ref{fil_2}(a).

Now each partner adds a filter at the end of the communication
channel. The most simple {\it commutative} filter one can consider
is convolution. Although this is a simple linear procedure, we
show that both hardware and software listeners fail to
synchronize.

The transmitted signal is now defined by
\begin{equation}
\label{Ttab} 
T_t^{A,B} = g_{A,B}(\vec{x}_t^{A,B}) = 
\sum_{\nu=0}^{N-1}{K_{A,B}^{\nu}f(x^{A,B}_{t-\nu})}
\end{equation}
where $K_A^{\nu},K_B^{\nu}\in[0,1]$ are the private keys
(filters) chosen randomly by each one of the partners, and
$\nu=0,1,\ldots,N-1$. We demand that
$\sum_{\nu=0}^{N-1}{K_{A,B}^{\nu}}=1$, in order to ensure that the
transmitted signal is limited by $[0,1]$. Before arriving to the
other end of the channel, the transmitted signal $T$ encounters the
second filter and the received signal is
\begin{eqnarray}
\label{Ttab2}
R_t^{A,B} = g_{B,A}(\vec{T}_t^{A,B}) = \sum_{\mu,\nu=0}^{N-1}{K_B^{\nu}K_A^{\mu}f(x_{t-\nu-\mu}^{A,B})} .
\end{eqnarray}
These quantities $R_t^{A,B}$ drive the units B and A, respectively.
For the case of $N=2$, equation (\ref{Ttab2}) yields
\begin{equation}
R_t^{A,B}=\alpha f(x_t^{A,B})+\beta f(x_{t-1}^{A,B})+\gamma
f(x_{t-2}^{A,B})\label{eqn_Tt}\end{equation} Where $\alpha=K_AK_B$,
$\beta=K_A(1-K_B)+K_B(1-K_A)$, and $\gamma=(1-K_A)(1-K_B)$. For such
a configuration the equations of dynamics are given by
\begin{equation}
\begin{split}
x_{t+1}^A=(1-\varepsilon)f(x_t^A)+\varepsilon\kappa
f(x_{t-\tau}^A)+\varepsilon(1-\kappa)R_{t-\tau}^B \\
x_{t+1}^B=(1-\varepsilon)f(x_t^B)+\varepsilon\kappa
f(x_{t-\tau}^B)+\varepsilon(1-\kappa)R_{t-\tau}^A
\end{split}
\end{equation}
Similarly to the case without filters, we calculate an expression for
the largest Lyapunov exponent and examine the regime of
synchronization. Since the values of the private keys $K_A,K_B$
are random, we calculate the probability of achieving
synchronization in the phase space of $(\varepsilon,\kappa)$ 
by sampling random sets of keys.
In figure \ref{fil_2} we compare
the regimes of synchronization for the basic setup with the lack
of filters (a) and static-filters setup (b). We found that even in
this case, the regime of synchronization is almost unchanged, and
in addition, there exists a large fraction of the phase space
where synchronization is achieved with a probability close to $1$.

\begin{figure}[h]
\begin{center}
\includegraphics[width=0.22\textwidth]{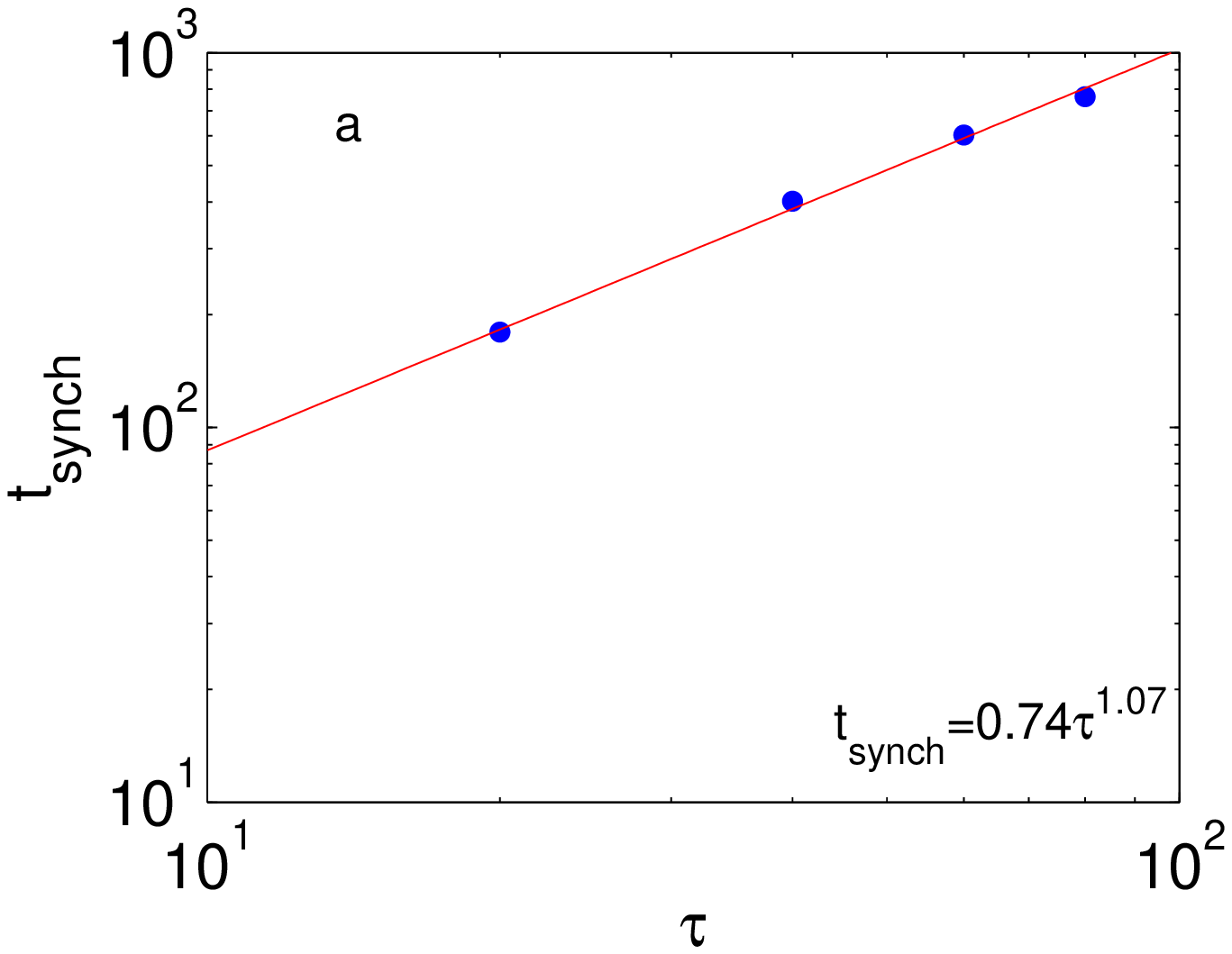}
\includegraphics[width=0.22\textwidth]{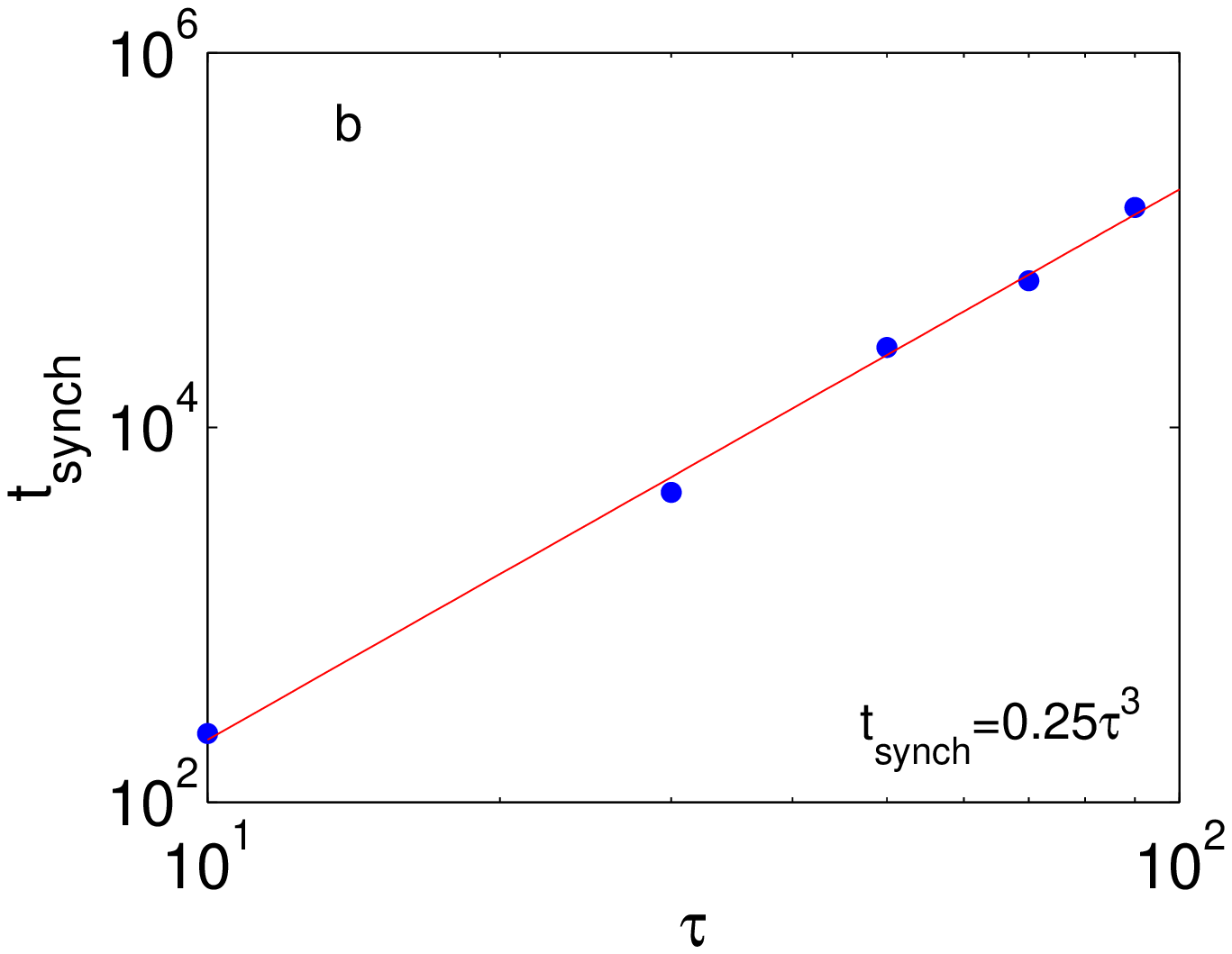}
 \caption{\label{fil_3}Synchronization time vs. $\tau$ (a) $K_A,K_B\in(0.89,0.91)$
 (b) $K_A,K_B\in(0.59,0.61)$.  The solid lines were obtained by linear regression indicating
 $\xi\sim 1.07,~3$ for (a) and (b), respectively.}
\end{center}
\end{figure}

The synchronization time, $t_{synch}$, is determined by simulations
and is found to scale with the parameters $(\tau,K_A,K_B)$ as
\begin{equation}
t_{synch} \propto \tau^{\xi(K_A,K_B)} .
\end{equation}
High values of keys, $K_A,K_B \rightarrow 1$, correspond to the case
of no filters, ($\alpha\rightarrow 1$ and $\beta,\gamma\rightarrow
0$), and ensure fast synchronization that grows almost linearly with
$\tau$, $\xi(K_A,K_B)\sim 1$. As we decrease the values of the keys
we find that  $\xi(K_A,K_B)$ increases, indicating a longer
synchronization time. In figure \ref{fil_3} we present $t_{synch}$
as a function of $\tau$ for different regimes of $(K_A,K_B)$. In
figure \ref{fil_3}(a), $K_A,K_B\in(0.89,0.91)$ therefore we find an
almost linear linear dependence $\xi \simeq 1.07$. In figure
\ref{fil_3}(b), $K_A,K_B\in(0.59,0.61)$ and we find a much slower
synchronization time, $\xi\simeq 3$.

\begin{figure}[h]
\begin{center}
\includegraphics[width=0.25\textwidth]{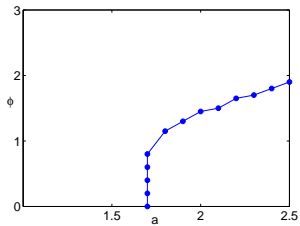}
\caption{\label{fil_5a}
Simulation results for $N=35$ indicate that synchronization in
$(a,\phi)$ is achieved only above the line. Similar results were
obtained also for larger values of $N$.}
\end{center}
\end{figure}

To ensure synchronization  for $N\gg 1$ we found that the 
strengths of the filter coefficients
have to follow a power-law
\begin{equation}
K_A^\nu,K_B^\nu\propto\frac{C^\nu_{A,B}}{\nu^\phi}
\label{eqn_keys}
\end{equation}
where $C^\nu_{A,B}$ is a random number between $[0,1]$,
and the dynamic equations are
\begin{multline}
x_{t+1}^A=(1-\varepsilon)f(x_t^A)+\varepsilon\kappa
f(x_{t-\tau}^A) + {}\\
{} + \varepsilon(1-\kappa)\sum_{\mu,\nu=0}^{N-1}{K_B^{\nu}K_A^{\mu}f(x_{t-\nu-\mu}^{B})}
\end{multline}
and similarly for $x_{t+1}^B$. 
The
largest eigenvalue can be found only semi-analytically, by
assuming that the distance between the partners converges/diverges
exponentially with time and then solving the characteristic
polynomial numerically. Results indicate that the fraction of the
phase space, $(\varepsilon,\kappa)$, where synchronization is
achieved does not alter as a function of the length of the key,
$N$, as long as the keys decay as a power-law, Eq.~(\ref{eqn_keys}), and $\phi$ is large enough (see figure
\ref{fil_5a}). However, the fraction of the phase space where
synchronization is achieved is strongly dependent on the interplay
between the two parameters of the coupled Bernoulli maps, $a$ and
$\phi$, see figure \ref{fil_5a}. The semi-analytic solutions
confirmed by simulation results indicate that the interplay
between $a$ and $\phi$ where synchronization is achieved can be
described as follows. For $a<a_c\sim 1.7$, synchronization is
independent of $\phi$. For $a\geq a_c$, synchronization is
achieved only above a critical value $\phi_c(a)$. 

We now turn to discuss the capabilities of different types of
listeners coupled unidirectionally. A hardware listener is
eavesdropping on the transmitted signals from both partners and in
order to synchronize he must imitate one of the filters,
$\{K_A^{\mu}\}$ or $\{K_B^{\mu}\}$. Since these  values are private
keys of each of the partners, he will not be able to recover the
received signal and to synchronize. 

Thus, only software listeners might be successful. By recording
the transmitted signal on both directions the listeners can
collect information and analyze the data in order to find first
the private keys. Once the keys are discovered, he can use them in
order to synchronize with the partners following the strategy
proposed in Eq.~(\ref{eqn_et}).

We first demonstrate a simple algebraic way of calculating
$K_A^\nu,K_B^\nu$ for $N=2$, 
assuming the partners are already
synchronized, 
$\vec{x}_t^A = \vec{x}_t^B \equiv \vec{x}_t$. 
In one time step, the transmitted signals on both
directions, 
$T_t^{A,B}=K_{A,B}f(x_t)+(1-K_{A,B})f(x_{t-1})$,
consists of four unknown variables: $K_{A,B},f(x_t),f(x_{t-1})$. On
the next time step, two new equations emerge:
$T_{t+1}^{A,B}=K_{A,B}f(x_{t+1})+(1-K_{A,B})f(x_{t})$. These
equations consist of previously three unknown variables
$K_{A,B},f(x_t)$ and one new unknown variable $f(x_{t+1})$.
Therefore by adding more time steps we are adding more equations
than new variables. Actually for $N=2$, three time steps supply $6$
equations for the $6$ unknown variables, and the keys, $K_{A,B}$,
can be revealed.

For $N>2$ the number of required equations to decode the keys of
length $N$ is $6(N-1)$. Note that the obtained equations are
nonlinear, but at least for $N=2$ and $3$ we found that the
solution is unique. In case the solution is unique also for $N>3$,
an unproven result, the complexity of finding the keys is similar
to the complexity of solving $6(N-1)$ linear equations.


In order to avoid the synchronization of a software listener we
propose to use \emph{time-dependent} filters. We replace the private keys
every few time steps, while the synchronization of the partners is
not damaged. This technique of key-swapping is exemplified below
for $N=2$. A necessary condition to maintain the synchronization
process is that the partners use commutative filters. Hence, the
filters cannot be changed every iteration, since the input signals
consist of the mixed output signals of three sequential time
steps, Eq.~(\ref{eqn_Tt}).

We suggest to use the same keys for two successive iterations and
stop transmission in the third iteration. In the third step we have
to increase the self-feedback to its maximum $\kappa=1$ to
compensate for the missing external input.
\begin{equation}
\begin{split}
 T_t^{A,B}=K_{A,B}^1f(x_t^{A,B})+(1-K_{A,B}^1)f(x_{t-1}^{A,B}) \\
 T_{t+1}^{A,B}=K_{A,B}^1f(x_{t+1}^{A,B})+(1-K_{A,B}^1)f(x_{t}^{A,B}) \\
 T_{t+3}^{A,B}=K_{A,B}^2f(x_{t+3}^{A,B})+(1-K_{A,B}^2)f(x_{t+2}^{A,B}) \\
 T_{t+4}^{A,B}=K_{A,B}^2f(x_{t+4}^{A,B})+(1-K_{A,B}^2)f(x_{t+3}^{A,B}) \\
\end{split}
\end{equation}
where $T_{t+2}^{A,B}=0$. We found stable synchronization
for this three steps protocol, where in each period new random
keys are selected. The regime of synchronization
$(\varepsilon,\kappa)$ was found to be similar to the case with
the absence of filters.

Generalizing the time-dependent random filters for $N>2$ requires
that the rate of the keys-swapping has to be greater than $2$ and
less than $6(N-1)$ iterations for communication, and at least
$N-1$ iterations with no communication, $\kappa=1$ \cite{note}, in
order to change the keys. As the key swapping consists of
$m<6(N-1)$ iterations, in each periodicity of the protocol the
listener is left with $6(N-1)-m$ unrevealed parameters of the key.
Simulations of the swapping protocol for $N$ up to few hundreds,
indicate that synchronization is achieved similarly to the case
with the absence of filters.

Note that using time-dependent filters eliminates any
reconstruction based on Taken's theorem, since the transmitted
signal is a discontinuous function of the chaotic variables, as
was suggested in \cite{Einat1}.


Our results show that indeed two way is better than one way, and
mutual is superior to unidirectional coupling. This superiority is
guaranteed even for the simplest chaotic maps and linear filters.
An attractive possible application based on this  phenomenon is
public channel encryption protocols. Although a listener knows the
parameters of the chaotic dynamics of the partners and the
transmitted signals he 
cannot reveal the private filters and cannot synchronize.

In the last few years, there were attempts to use coupled chaotic
lasers \cite{lasers5,lasers6,lasers7,nature} and coupled neural
networks \cite{nn1,nn2} for cryptography. Unidirectional coupling
was used as a private-key system, and mutual coupling for the
construction of public-channel cryptography.
\cite{lasers1,lasers2_MCPF,lasers3}.

The general proof that there is no advantage of mutual over
unidirectional coupling in the case of no filters, Eqs.\ (\ref{eqn_2dt})--(\ref{eqn_ct}), is in
question in the case of mutually coupled lasers.  It is not obvious
that enhancing the signal can be achieved without adding unavoidable
noise, moreover subtracting a signal (negative self-coupling in Eq.~(\ref{eqn_et})) is in question if possible at all. The subtraction
of a signal from itself requires that the lasers will be
synchronized not only with their amplitudes but also with their
optical phase. It is yet unknown if the optical phase is
synchronized in mutually or unidirectionally coupled chaotic lasers
and its accuracy. Hence there is still the possibility that even for
lasers with static filters,  the answer to our initial question
raised above  is positive. Note also that even with the lack of
filters, a public channel protocol for synchronized chaotic lasers
is possible. The messages are modulated on the chaotic transmitted
signals both way and recovered based on another phenomenon known as
a mutual chaos pass filter \cite{lasers2_MCPF}.

The research is supported in part by the Israel Science
Foundation.

\bibliography{dynamicfilters}

\begin{thebibliography}{20}
\bibitem[{not()}]{note}
\bibinfo{note}{Note that one can avoid the steps with no communication in case
  that each partner is dynamically changing the keys following a public
  function of his history $\vec{x}^{A,B}$.}
\end{thebibliography}

\end{document}